\documentclass[iop]{emulateapj}

\usepackage[applemac]{inputenc}
\usepackage[T1]{fontenc}
\usepackage[english]{babel}
\usepackage{graphicx}

\usepackage{amsmath}

\shorttitle{Alpha particle fire hose instability}
\shortauthors{Matteini et al.}

\begin{document}


\title{Fire Hose instability driven by alpha particle temperature anisotropy}


\author{L. Matteini\altaffilmark{1}, P. Hellinger\altaffilmark{2}, S. J. Schwartz\altaffilmark{1}, and S. Landi\altaffilmark{3}}
\altaffiltext{1}{Department of Physics, Imperial College London, SW7 2AZ London, UK}
\altaffiltext{2}{Astronomical Institute, CAS, Prague, Czech Republic }
\altaffiltext{3}{Dipartimento di Fisica e Astronomia, Università di Firenze, Italy}


\begin{abstract}
We investigate properties of a solar wind-like plasma including a secondary alpha particle population exhibiting a parallel temperature anisotropy with respect to the background magnetic field, using linear and quasi-linear predictions and by means of one-dimensional hybrid simulations. We show that anisotropic alpha particles can drive a parallel fire hose instability analogous to that generated by protons, but that, remarkably, the instability can be triggered also when the parallel plasma beta of alpha particles is below unity. The wave activity generated by the alpha anisotropy affects the evolution of the more abundant protons, leading to their anisotropic heating. When both ion species have sufficient parallel anisotropies both of them can drive the instability, and we observe generation of two distinct peaks in the spectra of the fluctuations, with longer wavelengths associated to alphas and shorter ones to protons. If a non-zero relative drift is present, the unstable modes propagate preferentially in the direction of the drift associated with the unstable species. The generated waves scatter particles and reduce their temperature anisotropy to marginally stable state, and, moreover, they significantly reduce the relative drift between the two ion populations. The coexistence of modes excited by both species leads to saturation of the plasma in distinct regions of the beta/anisotropy parameter space for protons and alpha particles, in good agreement with in situ solar wind observations. Our results confirm that fire hose instabilities are likely at work in the solar wind and limit the anisotropy of different ion species in the plasma.
\end{abstract}



\section{introduction}
Temperature anisotropies with respect to the direction of the local magnetic field are ubiquitous in the weakly collisional heliospheric plasma; particle distribution functions measured in situ by spacecraft at various heliocentric distances in the solar wind are not at thermal equilibrium and departures from Maxwellian are frequently observed \citep[e.g.,][]{Marsch_al_1982a, Neugebauer_al_1996, Gary_al_2006, Matteini_al_2013}.
The energy transport in non-thermal plasmas is an important topic for space and astrophysical environments, where microphysical processes generated by temperature anisotropies can importantly affect and change the macroscopic evolution of the system \citep{Matteini_al_2011}. 
Instabilities driven by a temperature anisotropy are thought to be an important mechanism in weakly collisional or collisionless plasma, where Coulomb collisions are not able to maintain the thermodynamical equilibrium of the gas and other processes take place to regulate the evolution of the particle distribution functions.
Direct measurements of the ion properties in the solar wind \citep[e.g.,][]{Hellinger_al_2006, Matteini_al_2007} and in the magnetosphere \citep{Samsonov_al_2007} suggest that kinetic instabilities driven by an ion temperature anisotropy play a role in constraining the ratio of parallel and perpendicular temperatures, defined with respect to the local mean magnetic field. Moreover, observations of electromagnetic spectra in the solar wind show some possible signatures of the presence of fluctuations associated to those processes \citep{Bale_al_2009, Wicks_al_2012}, although their interpretation needs some caution \citep[cfr.][]{Hellinger_Travnicek_2014}.
Kinetic instabilities like mirror, ion-cyclotron, and fire hose are also likely to be active in astrophysical systems, e.g. accretion disks \citep{Sharma_al_2006, Kunz_al_2014, Riquelme_al_2015, Sironi_Narayan_2015} and the hot plasma in galaxy clusters \citep{Schekochihin_al_2005, Mogavero_Schekochihin_2014}, and in plasmas where magnetic reconnection is at work \citep{Drake_al_2010, Schoeffler_al_2012, Matteini_al_2013b, Gingell_al_2015}.

Among the different instabilities, fire hose is expected to be the dominant process in regulating the temperature anisotropy of a weakly collisional plasma when the parallel temperature is larger than the perpendicular temperature.
Note that this condition is naturally obtained in systems where the magnetic field is spatially modulated and decreasing with distance (as for example expanding plasmas, like the solar and other magnetized stellar winds) due to particle magnetic moment conservation \citep{Matteini_al_2011}
or where field aligned beams - contributing to an excess of the total parallel  temperature - are generated, as typically observed in reconnection exhausts \citep[][Hietala et al., submitted]{Hoshino_al_1998, Gosling_al_2005}.

Previous studies of fire hose instabilities going beyond the fluid approximation have focussed on both electrons \citep[e.g.,][]{Li_Habbal_2000, Messmer_2002, Paesold_Benz_2003, Camporeale_Burgess_2008, Hellinger_al_2014} and protons \citep[e.g.,][]{Quest_Shapiro_1996, Gary_al_1998, Hellinger_al_2003, Matteini_al_2006, Hellinger_Travnicek_2008}.
These studies are supported by \emph{in situ} observations in the solar wind, for protons \citep{Kasper_al_2002, Hellinger_al_2006} and electrons \citep{Stverak_al_2008}.
However, much less attention has been addressed, to our knowledge, to the case when the fire hose instability is driven by other species of the plasma.
While the alpha particle abundance in the solar wind is only few percent, they are heavier and usually both faster and hotter than protons so that they carry important fractions of the solar wind momentum and energy. Their deceleration with respect to
the protons may serve as an important source of energy for the proton heating \citep{Schwartz_Marsch_1983, Hellinger_Travnicek_2013, Verscharen_al_2015}.

Recent observations of the solar wind alpha particle distribution functions \citep{Maruca_al_2012}, have shown that temperature anisotropies of alpha particles also 
appear to be constrained by boundaries in plasma beta/anisotropy space, suggesting the role of micro-instabilities in the regulation of alpha properties. Moreover, \cite{Bourouaine_al_2013} have shown that alpha-protons drifts may also be influenced by these processes.

How this happens and how different instabilities driven by different species simultaneously act in a plasma and stabilize particles to different marginally stable states has not been investigated in detail yet and it is the aim of this paper to do it by means of both linear and quasi-linear theoretical calculations and fully nonlinear numerical simulations.
We show results from 1-D numerical hybrid simulations that describe the development of the parallel fire hose, which is the dominant instability when a composition compatible with the solar wind plasma is adopted (see below), in the presence of anisotropic alphas with variable properties.
Nevertheless, results obtained in this particular framework may be applied, with a suitable change in the parameters adopted, to any other space or astrophysical plasma.

The paper is organized as follows: in Section \ref{linear} we present linear theory predictions of the parallel fire hose, focussing in particular on the differences between proton and alpha contributions in driving the instability and the influence of a velocity drift between the species; in Section \ref{quasilin} we discuss the feedback of the instability on the plasma, deriving the ion parallel and perpendicular heating rates and momentum exchange expected on the basis of a quasilinear approach; in Section \ref{results} we show results from hybrid simulations for different parameter combinations, including both the case of purely alpha driven fire hose as well as cases with mixed ion contributions. Finally in Section \ref{discussion} we summarize and discuss the results obtained, comparing linear and quasi-linear predictions with the fully nonlinear evolution found in simulations and in Section \ref{conclusion} we present our conclusion, highlighting implications of our study for solar wind and astrophysical plasmas.

\begin{figure}
   \centering
   \includegraphics[width=8.5cm]{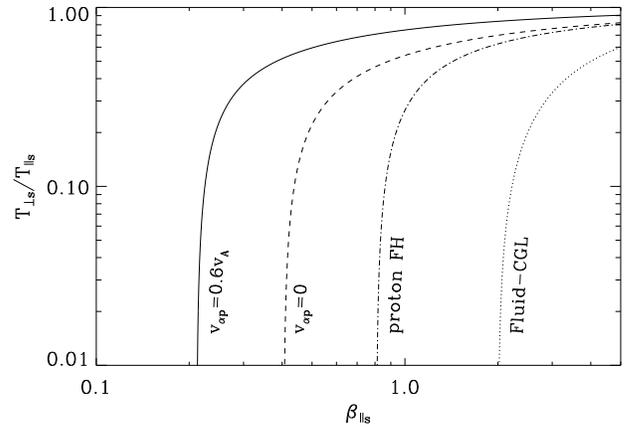} 
   \caption{Fire hose unstable region in the presence of alpha particle temperature anisotropy; 
   contours of the $\gamma_m=10^{-3}$ maximum growth rate level are shown for a case with an alpha-proton drift $v_{\alpha p}=0.6v_A$ (solid line), and with $v_{\alpha p}=0$ (dashed line). The proton driven ($n_\alpha=0$) and fluid CGL fire hose thresholds are shown as references in dash-dotted and dotted lines, respectively.}
   \label{fig_region}
\end{figure}

\begin{figure*}
   \centering
   \includegraphics[width=13cm]{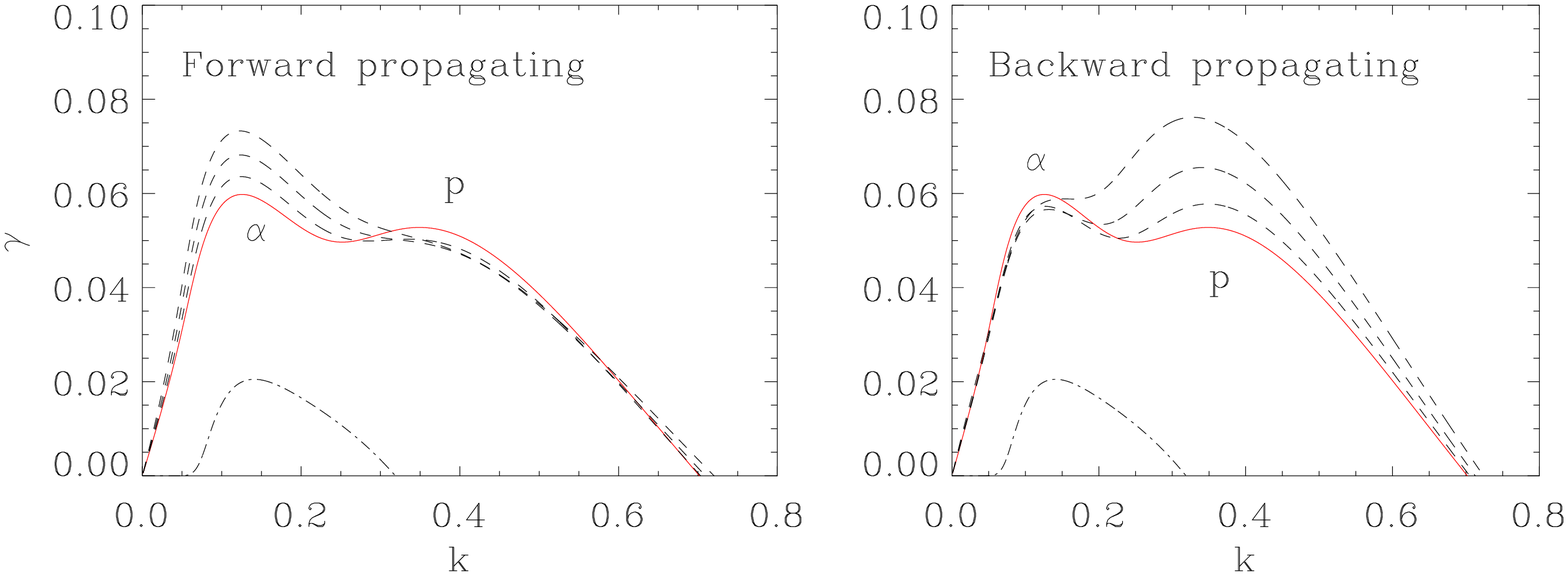} 
   \caption{Instability growth rates for $\beta_p=3$ and $\beta_{\alpha}=2$ and anisotropic protons $A_p=0.5$ and alphas $A_\alpha=0.4$. Left panel refers to forward propagating waves and right panel to  backward propagating modes.
   The red line shows case with $v_{\alpha p}=0$, while dashed line refers to different values of the alpha-proton drift $v_{\alpha p}=0.3, 0.6, 0.9 v_A$. The dash-dotted line shows the growth rate for the same alpha parameters but isotropic protons.}
   \label{fig_gamma}
\end{figure*}

\section{Linear theory}\label{linear}
We start our analysis with linear predictions, obtained solving the dispersion of a fully kinetic Vlasov system.
The dispersion and stability of the plasma depends on many parameters and the presence of three populations complicates further the picture with respect to the case of a pure electron-proton plasma. 
For a detailed linear theory investigation of the role of alpha particles and their influence on the dispersion see for example \cite{Maruca_2012}.
Two fire hose instabilities are in general active in plasmas with $T_\perp<T_\|$ (see Appendix  for symbol definitions): the parallel fire hose \citep[e.g.,][]{Quest_Shapiro_1996, Gary_al_1998}, with maximum growth rate at parallel propagation, and the oblique fire hose \citep{Hellinger_Matsumoto_2000, Hellinger_Matsumoto_2001}, which has a maximum for oblique wave-vectors. In a plasma with proton anisotropy the parallel instability is typically dominant for $\beta_{\| p}\lesssim10$.  
The same behavior is predicted by the linear theory when the instability is driven by temperature anisotropy in alpha particles, with a preferential excitation of fluctuations parallel to the magnetic field. For this reason, in this work we specialize to the case of the dominant parallel fire hose instability.

Figure \ref{fig_region} shows the fire hose unstable region in the parameter space ($\beta_{\| s},T_{\perp s}/T_{\| s}$) derived from Vlasov linear theory for different plasma configurations.
Contours display the $\gamma_m=10^{-3}$ maximum growth rate level, which can be reasonably taken as an estimation of the threshold (or rather a marginal stability condition) of the instability \citep[e.g.,][]{Matteini_al_2006}.
Solid and dashed lines are associated with anisotropic bi-Maxwellian alpha particles, with density $n_\alpha=0.05n_e$; for these computations (where the axis refers to alpha particles) protons and electrons are isotropic, and $\beta_{p}=\beta_{e}=1$.
The solid line shows the maximum growth rate of the alpha fire hose instability when an alpha-proton drift $v_{\alpha p}=0.6v_A$ is assumed in the plasma; note that this is a typical value in the fast solar wind \citep[e.g.,][]{Neugebauer_al_1996}.
The dashed line refers to the same level for the case with $v_{\alpha p}=0$. 
As a reference, the dotted line shows the CGL-fluid fire hose threshold: $\beta_\|-\beta_\perp=2$.
Note first that the unstable region predicted for anisotropic alphas extends to much smaller betas with respect to the fluid case. 
This remains true also when smaller values are assumed for the proton and electron betas, which contribute to the total plasma pressures, underlining that anisotropic alpha particles can drive a kinetic fire hose instability also when the total plasma beta is $\ll2$, thus entirely stable with respect to the corresponding fluid instability. Such a linear prediction is confirmed by numerical simulations, as we will show in what follows.
For the choice of parameters used here, the alpha-driven unstable region is also larger than the analogous proton parallel fire hose instability region predicted for typical solar wind conditions (see \citep{Matteini_al_2006}) when alphas are not taken into account, which is approximatively bounded by $\beta_{\| p}\gtrsim1$  and is shown as the dash-dotted contour.

Second, the unstable region is  wider in the case of a non-zero alpha-proton relative drift, and in general the growth rate significantly depends on the value of $v_{\alpha p}$.
Note that the presence of a relative drift between species also influences the spectrum of waves excited by the instability in terms of their direction of propagation, and then modulates the associated scattering of particles \citep[e.g.,][]{Verscharen_al_2013}.
In the absence of a velocity separation between protons and alphas a balance in the generation of forward and backward propagating modes is expected.
By contrast, in case of an alpha-proton relative drift, this symmetry is broken; if $v_{\alpha p}\ne0$ then dominant unstable waves of each species are expected to propagate in opposite directions.
Resonant fire hose is driven by the anomalous-doppler resonance \citep[e.g.][]{Kennel_Scarf_1968}: 
\begin{equation}\label{eq_resonance}
\omega-k_\|v_\|=-\Omega_s\,,
\end{equation}
where $\omega$ and $k_\|$ are respectively the frequency and the parallel wave-vector of the unstable wave, $v_\|$ is the parallel velocity of the resonant particles, and $\Omega_s$ is the cyclotron frequency of species $s$
(Note that in this resonance the Doppler-shifted frequency has opposite sign with respect to the standard cyclotron resonance: ${\omega-k_\|v_\|=\Omega_s}$).
Resonant parallel fire hose instability destabilizes right-handed magnetosonic/whistler waves, and since particles satisfy the resonant condition (\ref{eq_resonance}) with $kv_\|>0$, we expect then the unstable modes to be preferentially generated in the direction of the drift of the corresponding species.
For alphas drifting ahead protons along the magnetic field, this means they support mostly forward propagating waves, while protons tend to excite more backward propagating modes.

\begin{figure*}
   \centering
   \includegraphics[width=16.5cm]{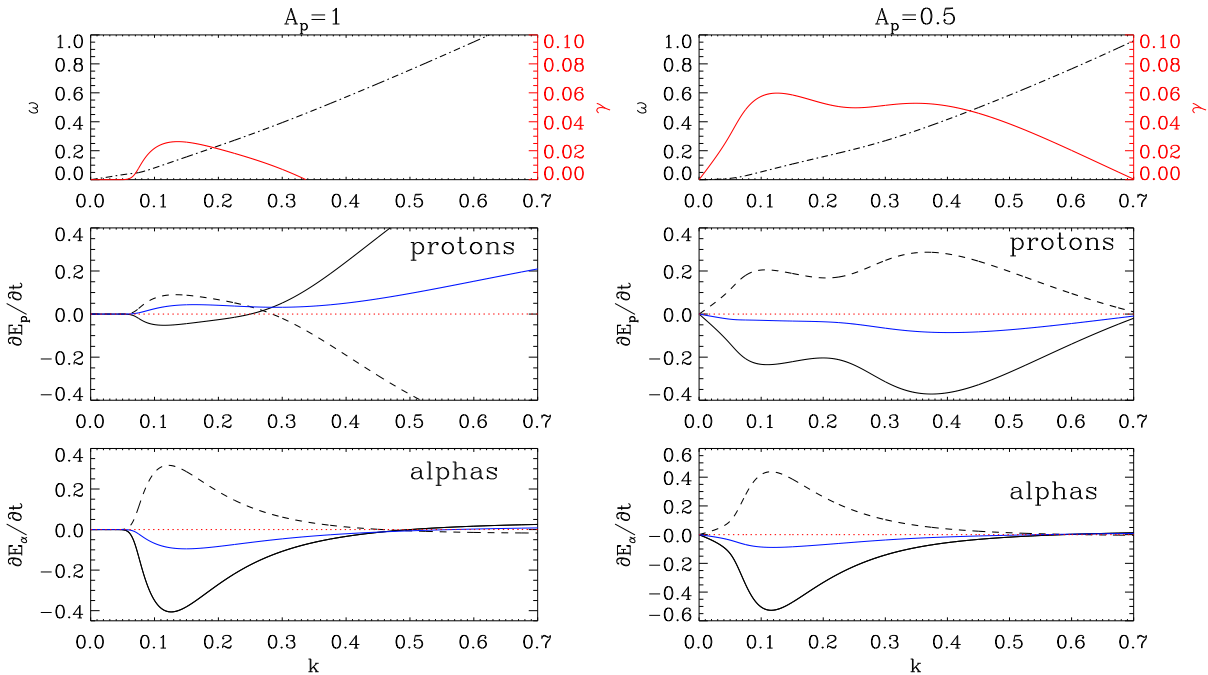} 
   \caption{Linear and quasi-linear predictions for the same plasma parameters as in Fig.~\ref{fig_gamma}: $A_\alpha=0.4, \beta_p=3$, $\beta_{\alpha}=2$, and with $A_p=1$ (left panels, dash-dotted line case in Fig.~\ref{fig_gamma}) and with $A_p=0.5$  (right panels, red line case in Fig.~\ref{fig_gamma}).
   Top: real frequency (dot-dashed) and growth rate (red) as a function of the unstable wave-vectors $k$.  Quasilinear predictions of heating rates $\partial E /\partial t$ as a function of $k$ for protons (middle) and alphas (bottom) are also shown; dashed and solid lines refers to parallel and perpendicular directions, respectively. The blue line shows the total heating rate and the dotted red line the total energy of the system. Heating rates are normalized following \cite{Hellinger_al_2013} and negative values correspond to particle cooling.}
   \label{fig_ql}
\end{figure*}

This scenario is confirmed by linear theory and it can be seen in Fig.~\ref{fig_gamma}, where the two panels show predictions for unstable forward (left panel) and backward (right panel) propagating modes.
The red line (same in both panels) shows the predicted growth rates in case when no drift is present and both species are anisotropic and unstable ($A_p=0.5$ and $A_\alpha$=0.4; $\beta_p=3$ and $\beta_{\alpha}=2$); in this case symmetry of the system produces equal amount of waves in both direction.
Also note that the presence of anisotropic protons significantly enhances the growth rate of the instability with respect to the case when only alphas are anisotropic, shown in dash-dotted line, and that this is true also in cases when protons are anisotropic but stable with respect to the fire hose.
It is interesting to observe that, for the parameters used in Figure \ref{fig_gamma}, there are two distinct peaks in the growth rate as a function of wave-vectors. These correspond to distinct fire hose modes driven by proton and alpha anisotropies, with the lower $k$ peak associated with the alphas and the other with the protons. This is a general feature characteristic of the regime investigated in this work, and is also confirmed by numerical simulations.

As expected when a relative drift is introduced ($v_{\alpha p}/v_A=0.3, 0.6$, and $0.9$, shown by dashed lines), the symmetry between left and right panels is broken. As $v_{\alpha p}$ is increased the peak corresponding to alpha fire hose is enhanced for forward waves (same direction as the drift  of alphas in the electron/plasma frame), while it is slightly weakened for backward waves. The opposite is observed for protons, for which the corresponding peak increases for backward waves and decreases for forward, owing to their negative drift in plasma frame.

\section{Quasilinear predictions}\label{quasilin}
A further step in the analysis can be done estimating quasilinear heating rates  associated to the linear predictions.
\cite{Hellinger_al_2013} estimated the initial quasilinear parallel and perpendicular heating rates (per mode) by applying the quasilinear diffusion operator (containing only the contribution of the given mode) on the initial bi-Maxwellian particle velocity distribution functions and calculating the corresponding moments.
Following this approach, we have then calculated the quasilinear plasma response to fluctuations generated by unstable ions, for parameters studied in Fig.~\ref{fig_gamma}: $A_\alpha=0.4, \beta_p=3, \beta_{\alpha}=2$, and as a function of the range of unstable modes $k$. 
Fig.~\ref{fig_ql} shows the real and imaginary part of the complex frequency from linear theory (top), and the proton (middle) and alpha (bottom) quasilinear predictions.
In the second and third panels, the dashed line shows the perpendicular heating rate, the solid line the parallel heating rate (negative values mean cooling), the blue line shows the total kinetic (thermal) energy and the red dotted line shows the total energy of the system (kinetic$+$magnetic) that is conserved to a very good approximation. Heating rates are normalized to $E_{em}\omega_{p}$, where $E_{em}$ is the electromagnetic energy of a single interacting mode (see \cite{Hellinger_al_2013} for details).

A case when only alpha particles are anisotropic and drive the instability - corresponding to the dash-dotted line in Fig.~\ref{fig_gamma} ($A_p=1$) - is shown in the left panels of Fig.~\ref{fig_ql}.
This calculation results in a perpendicular heating and a parallel cooling for alphas (middle panel), as expected for fire hose instability; such  an energy transfer leads to the isotropization of the initial $T_\perp<T_\|$ condition, owing the self-regulating behavior of the instability.
It is interesting to note that the same response is predicted also for protons (bottom-left panel); quasilinear approach predicts perpendicular heating and parallel cooling for the isotropic and stable protons, as a response to the generation of fluctuations by the alphas. As it will be discussed in the next session, this prediction is confirmed by fully nonlinear numerical simulations.
This also means that part of the free energy extracted from alpha particles is transferred to protons, via diffusion by the unstable fire hose modes. Note that this also leads to the generation of a proton temperature anisotropy $T_\perp>T_\|$.

The pictures is further complicated when both species are initially anisotropic and contribute to the instability. As discussed previously, linear theory predicts distinct unstable waves associated with the two populations. In this case particles can also resonate with modes driven from the other species.
This is shown in the right panels of Fig.~\ref{fig_ql} for the same parameter combination as the red line in Fig.~\ref{fig_gamma} ($A_p=0.5$ and $v_{\alpha p}=0$). By comparing the three panels, it can be noticed that each species is isotropised mostly in correspondence to the peak of the growth rate associated with their anisotropy. However, protons are significantly anisotropically heated/cooled also by fluctuations that are driven by alphas, at lower $k$. By contrast, modes generated by protons ($k\gtrsim 0.3$) have little influence on alphas.
 
If a drift velocity is present, then also a slowing down of the alpha-proton drift is expected, together with the change in the thermal energetics described above. 
The rate of the change in the kinetic energy associated with the proton-alpha drift induced by fire hose can be analogously derived from the quasilinear theory (Hellinger et al. 2013).
In this case, the quasilinear calculation predicts a decrease of the drift between the two species, with a fraction of the kinetic energy associated with the differential motion converted into thermal energy (and also into wave generation), leading to further anisotropic heating of both species.
These theoretical quasilinear predictions, however, get complicated when too many parameters are taken into account, and, moreover, they are only valid initially when the particle distribution functions are close to bi-Maxwellian. For a more complete description of the plasma evolution, numerical simulations are a more suitable tool.

\section{Simulations}\label{results}
We have performed one-dimensional (1-D) numerical simulations using a hybrid code \citep{Matthews_1994} which treats protons as particles and electrons as a massless charge neutralizing fluid with a constant temperature.
In the code units of space and time are the ion inertial length and 
the inverse proton cyclotron frequency, respectively. See the Appendix for definition of code units.
We use a simulation grid of 1200 collocation points, with spatial resolution $\Delta x=1$ and $10^4$ particles per cell for each species.

\begin{table}[b]\label{table}
\caption{Initial parameters for Hybrid 1-D simulations}
\begin{center}
\begin{tabular}{|c|c|c|c|c|c|c|c|c|}
\hline
Run & $\beta_{\| p}$ & $A_p$ & $\beta_{\| \alpha}$ & $A_\alpha$ & $v_{\alpha p}/v_A$ & $\beta_e$ & $\beta_{\| tot}$\\
\hline
A & 0.1 & 1 & 0.4 & 0.1& 0 & 0.01 & 0.41\\
B & 2 & 1 & 2 & 0.2 & 0& 1 &5\\ 
C & 6 & 1 & 2 & 0.2 & 0& 3 &11\\ 
\hline
D & 1.5 & 0.2 & 0.7 & 0.15 & 0& 0.5 & 2.7\\ 
E & 2 & 0.3 & 1 & 0.25 & 0& 0.5 & 3.5\\ 
F & 3 & 0.5 & 2 & 0.4 & 0& 1 &6\\ 
\hline
G & 1 & 1 & 0.5 & 0.5 & 0.9 & 0.5 & 2\\ 
H & 2.5 & 0.5 & 0.6 & 0.5 & 0.9& 0.5 &3.6\\
I & 3 & 0.5 & 2 & 0.4 & 0.9 & 1 & 6\\
 \hline
\end{tabular}
\end{center}
\label{default}
\end{table}%

\begin{figure}
   \centering
   \includegraphics[width=8.5cm]{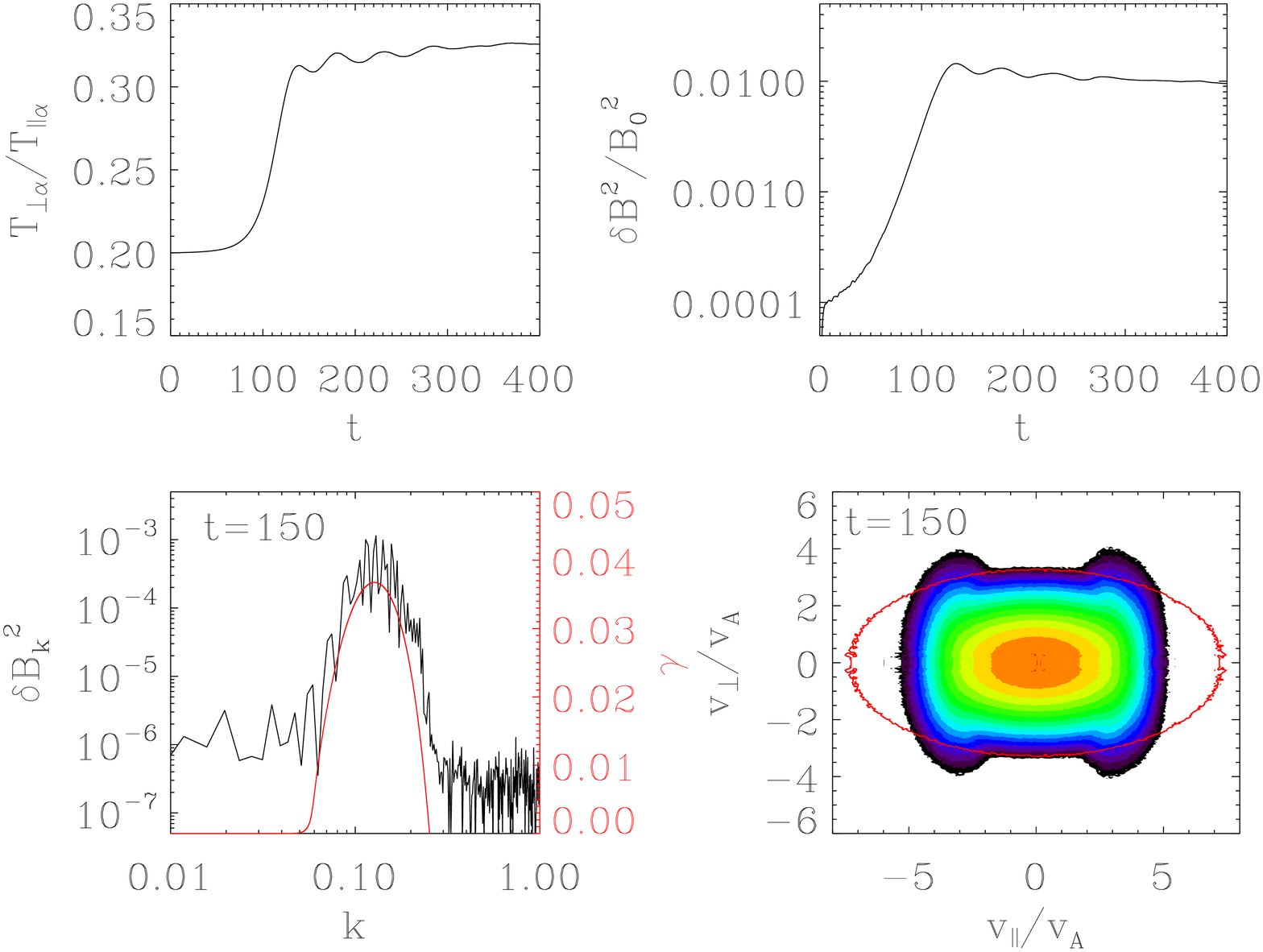} 
   \caption{Example of alpha fire hose instability (Run C). Top left: alpha temperature anisotropy evolution; Top right: magnetic energy evolution; Bottom left: spectrum of fluctuations at $t=150$ and linear theory prediction of instability growth rates (red); Bottom right: alpha particle distribution at $t=0$ (red line) and $t=150$.}
   \label{fig_example}
\end{figure}

We have performed several runs, with different initial conditions corresponding to the different cases discussed through the previous sessions and summarized in Table~\ref{table}. Run A-C focus on the case when only alpha particles are anisotropic over different beta regimes, Run D-F describe configurations where both protons and alphas are anisotropic, in order to study the coexistence of the modes driven by both ions, and finally Run G-I include a relative drift between the species.
The main results of these simulations are presented and discussed over the next sections.

\begin{figure}
   \centering
   \includegraphics[width=8.5cm]{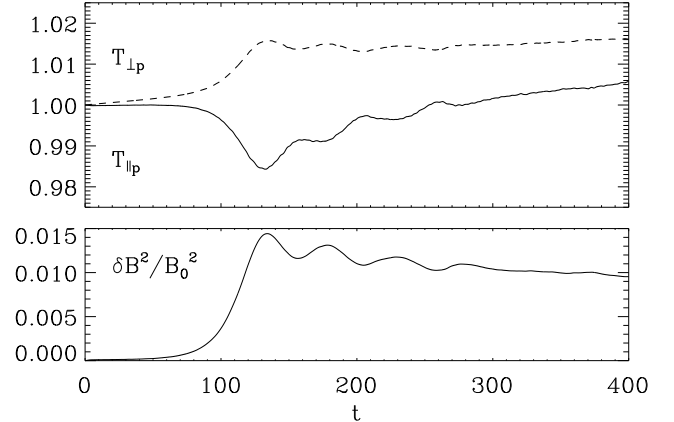} 
   \caption{(Top) Evolution of parallel and perpendicular proton temperatures in Run C; (bottom) evolution of the power of magnetic fluctuations as in Fig.~\ref{fig_example}}
   \label{fig_example_p}
\end{figure}

\subsection{Isotropic protons}
We start with a configuration where only alpha particles are initially anisotropic. Figure\,\ref{fig_example} provides an overview of the dynamics observed for Run C, which has plasma parameters consistent with solar wind conditions \citep{Maruca_al_2012}.
Top panels show the temporal evolution of the anisotropy (left) and of the magnetic fluctuations (right).
A linear phase of the instability is visible in the interval $50\lesssim t\lesssim100$, as suggested by the exponential growth of the magnetic fluctuating energy, which then saturates after $t\sim120$.
As the unstable fluctuations develop, they start scattering particles reducing their initial anisotropy. At saturation the plasma has reached a marginally stable state with $T_{\perp \alpha}/T_{\| \alpha}\sim0.4$.
The bottom left panel shows the spectrum of the fire hose fluctuations at $t=120$, at saturation. The red line corresponds to the growth rate predicted by the linear theory. There is a very good agreement between the range of unstable modes predicted by the theory and the spectrum observed in the simulation after saturation. Finally, in the right bottom panel the alpha particle distribution function after saturation is shown; the shape of the distribution at $t=0$ is also shown as a red contour.
As a consequence of the resonant wave-particle interaction between alpha particles and fire hose fluctuations, pitch-angle scattering signatures in velocity space are observed producing a significant deviation from the initial bi-Maxwellian.
The deformation of the distribution is similar to the one observed for protons in the case of the parallel \citep{Matteini_al_2006} and oblique fire hose \citep{Hellinger_Travnicek_2008}.
Note that, as discussed, the position of the unstable spectrum in k-space and the signatures in the particle distributions depend on resonances and are then a function of the specific plasma parameters taken as initial conditions; however the same qualitative global behavior of the instability shown in Fig. \ref{fig_example} is found also in simulations with different (lower) initial alpha and proton betas (Run A-B). 
Note that in Run A the total plasma beta is smaller than 1, corresponding then to a condition theoretically stable with respect the CGL-fluid fire hose (see Figure\,\ref{fig_region}). Although plasma parameters in this case are not realistic in terms of solar wind conditions, resulting in an excessively large alpha-proton temperature ratio, more generally Run A highlights the kinetic nature of the instability.

It is interesting to investigate the feedback on the background dominant proton population from the dynamics driven by the secondary alpha population.
As shown in the previous section, the quasilinear approximation predicts an anisotropic heating/cooling of the proton distribution by the modes excited by the alphas; in particular the generation of a proton temperature anisotropy is expected.
This is confirmed by Figure \ref{fig_example_p}, where the solid and dashed lines in the top panel refer to the evolution of the parallel and perpendicular proton temperatures, respectively. The development of a proton anisotropy is very well correlated to the growth of unstable fluctuations driven by the alpha fire hose instability (bottom panel). Moreover, the temperature evolution predicted by the quasilinear calculation (perpendicular heating and parallel cooling) describes well the evolution of the temperature profiles during the initial phase of the instability.
After saturation ($t\sim150$), well beyond the validity of the linear and quasilinear approximations, a parallel proton heating is observed, which progressively reduces the proton anisotropy developed during the linear phase of the instability; this is likely related to the late damping of the fire hose fluctuations, as a slight decrease in the magnetic field power after saturation is observed (bottom panel).

\subsection{Alphas and protons unstable}
\begin{figure}
   \centering
   \includegraphics[width=9cm]{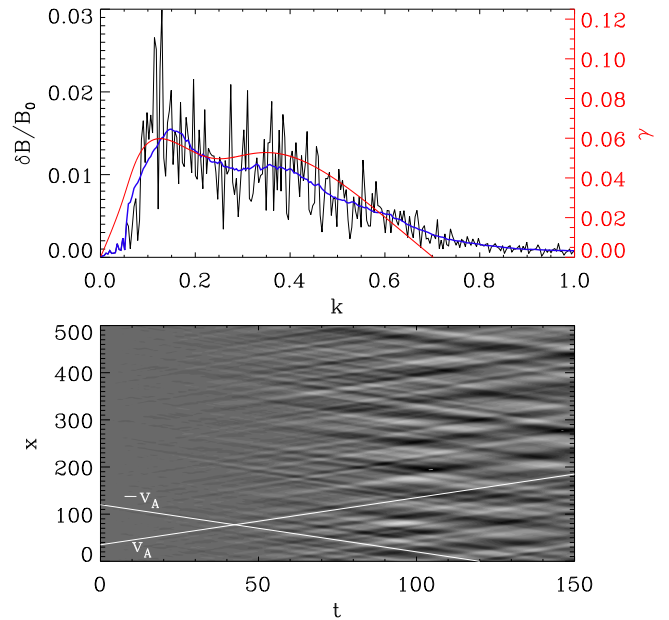} 
   \caption{Run F: (top) Spectrum of the unstable modes at $t=80$. Red line shows the linear prediction for the same parameters as in the run (See Fig.~\ref{fig_gamma}), while the blue line is a smoothed profile of the simulation data. (bottom) Profile of the $B_z$ component of the magnetic field as a function of time. The white straight lines show the propagation of forward and backward waves with phase speeds $\pm v_A$.}
   \label{fig_gamma_run}
\end{figure}

We consider now the situation when both species are initially anisotropic, which constitutes a more realistic description of the solar wind regime and possibly of weakly collisional astrophysical plasmas in general. 
In this configuration, when starting with both species predicted unstable by the linear theory, we expect to observe a co-existence of the proton and alpha fire hose instabilities.
Figure \ref{fig_gamma_run}, top panel, shows the spectrum of fluctuations for Run F, for which a double-peaked spectrum of fluctuations, as in Fig. \ref{fig_gamma}, is expected. The spectrum observed in the simulation is in good agreement with the linear prediction of the unstable modes (red line), showing distinct contribution from proton and alpha unstable modes. The blue line shows a smoothed profile of the spectrum in the simulation, and highlights the presence of a double peaked profile.
As discussed in section \ref{linear}, the generation of unstable fluctuations at larger wavelength is associated with the modes driven by the alphas (consistent with the lower $k$ single peaked spectrum observed in Fig.~\ref{fig_example}, when protons are stable), while smaller wavelengths correspond to proton modes. Moreover, distributions are symmetric with respect to the magnetic field direction, so that modes propagating in both directions are developed in the simulation, as expected; the bottom panel of the figure shows a profile of the magnetic field component $Bz$ as a function of time, highlighting the presence of counter-propagating wavefronts approximatively propagating at $\pm v_A$ (white trajectories in solid line), driven by the instabilities after $t\sim40$.

\begin{figure}
   \centering
   \includegraphics[width=8.5cm]{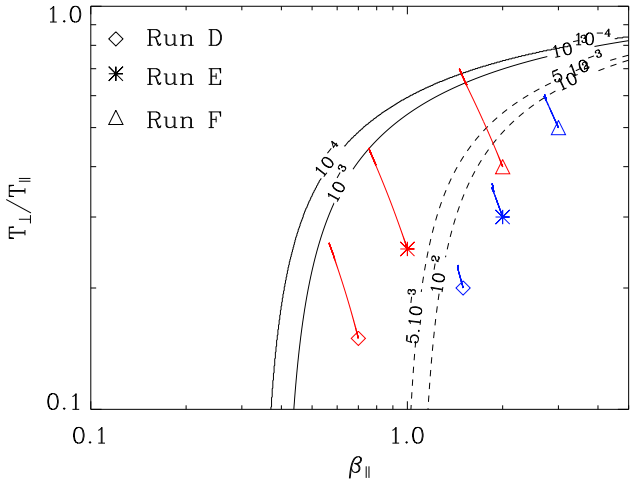} 
   \caption{Evolution of anisotropies for alphas (red) and proton (blue); symbols show the unstable initial conditions for 3 different simulations:  Run D (diamonds), Run E (stars), and Run F (triangles). The dashed and solid contour lines show the constant growth rates levels of proton and alpha parallel fire hoses predicted by the linear theory, respectively.}
   \label{fig_ev_anis1}
\end{figure}

In order to investigate further the evolution of the instability when the two species are both unstable, it is useful to use the $(\beta_{\| s},T_{\perp s}/T_{\| s})$ plane.
Figure \ref{fig_ev_anis1} shows trajectories in this parameter space for three simulations with different initial conditions (Run D-E-F: solid, dashed and dotted line, respectively).
Note that parameters of these runs, chosen in order to emphasize the simultaneous evolution of both species close to fire hose instability thresholds, lead to an initial alpha-proton temperature ratio which is larger than typical solar wind conditions. However, we have verified that this choice does not change significantly the results and does not affect the qualitative evolution of the plasma with respect to simulations with a more realistic temperature ratio.
The solid contours correspond to the alpha fire hose unstable region, as in Figure \ref{fig_region}, while the dashed contours show the instability threshold for the instability driven by protons. The red and blue lines indicate the evolution path of alphas and protons, respectively, and symbols show the initial conditions. Both species are initially unstable and then become more isotropic during the simulations; this corresponds to a migration towards the top-left corner in the parameter space. 
Remarkably, each species is stabilized according to its own marginal stability, thus in different regions of the parameter space: protons saturate at $\beta_{\| p}>1$, while alphas are pushed towards their threshold (solid contours) that extends also to $\beta_{\| \alpha}<1$.
The final stable state found after the nonlinear stage in the simulation is in reasonable agreement with marginal stability given by linear theory. Note however that due to the resonant interaction between particles and unstable modes, distributions may deviate significantly from Maxwellian at saturation; as discussed in \cite{Matteini_al_2006}, this can lead to the fact that the plasma saturates in the parameter space before  reaching the theoretical linear thresholds that are computed for bi-Maxwellian distributions. It is then not surprising that the agreement between simulations and linear theory in Fig.~\ref{fig_ev_anis1} is not exact, although it captures the global trend of the stabilization.

\subsection{Role of alpha-proton drift}
We finally consider the role of an alpha-proton drift in the dynamics of the instability.
As shown in section \ref{linear},  the presence of a relative streaming between species introduces an asymmetry in the system, owing to each species' drift with respect to the plasma (electron fluid) frame.
As a consequence, we expect to observe in the simulation a preferred direction of propagation in the modes driven by each species, following the direction of their drift in the plasma frame.
To confirm this scenario, we present in this section results from runs where an initial drift of the order of the Alfvén speed is taken into account.
This is a characteristic often observed in the fast solar wind \citep{Marsch_al_1982b, Neugebauer_al_1996, Goldstein_al_2000}.

\begin{figure}
   \centering
   \includegraphics[width=8.5cm]{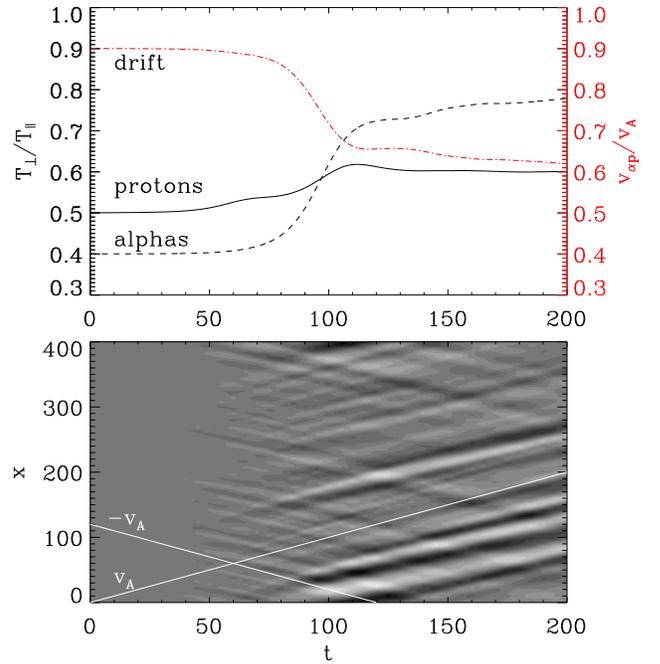} 
   \caption{Fire hose instability in the case of an alpha-proton drift $v_{\alpha p}=0.6v_A$ (Run I); in the top panel the solid line shows the alpha anisotropy, while dashed line refers to proton anisotropy, and red dot-dashed to the alpha-proton drift. In bottom panel, the profile of the transverse magnetic filed $B_z(x)$ as a function of time is shown. The solid lines corresponds to the trajectory of wave propagating with $\pm v_A$.}
   \label{fig_run2}
\end{figure}

Figure \ref{fig_run2} shows the results for Run I, which has the same initial conditions as Run F, but with ${v_{\alpha p}=0.9v_A}$.
The top panel shows the evolution of temperature anisotropies; protons are affected first, at $t\sim30$, leading to a small change in their anisotropy, but sufficient to stabilize them. The activity driven by the protons has little influence on the alpha anisotropy, consistent with the quasi-linear prediction of Fig.~\ref{fig_ql}.
Shortly after, from $t\sim50$-$60$, alpha particles are isotropized more significantly, corresponding to the phase when fluctuations associated with the alphas start to play a role.
A second change in the proton anisotropy (solid line), is also observed at $t\sim80$, during the isotropization of alphas. This phase is analogous to the anisotropic proton heating seen in  Figure \ref{fig_example_p} (only alphas unstable) and it is also consistent with the quasilinear prediction discussed in section \ref{quasilin} and shown in Figure \ref{fig_ql}.

During the activity of the instability the alpha-proton relative drift is also influenced. The red dash-dotted line in the top panel shows the evolution of $v_{\alpha p}$ as a function of time. A first weak decrease can be observed when protons become unstable, while the largest variation occurs later, when the alpha fire hose instability takes place; at the end of the run, the drift between the two ion population is reduced from $v_{\alpha p}=0.9v_A$ to $v_{\alpha p}=0.6v_A$. 
We have checked that this behavior, which is in agreement with the quasilinear predictions, is confirmed by simulations where protons are isotropic (Run G) and with $\beta_{\| \alpha} <1$ (Run H).

The lower panel of Fig.~\ref{fig_run2} shows the profile of the transverse magnetic field component $B_z(x)$, inside the region $x=[0,400]$ and as a function of time. 
The white solid lines correspond to the trajectory of waves propagating with phase velocity $\pm v_A$.
Remember that, according to Figure \ref{fig_gamma}, the linear theory predicts for protons a  dominance of backward propagating fluctuations, while alphas are expected to excite mainly forward propagating waves. 
Consistent with the expectations and  the evolution of the anisotropies shown in the top panel, we observe the presence of backward waves during the first phase of the run, when the instability is firstly driven by protons, due to their initial negative drift. 
After $t\sim70$, when the macroscopic dynamics begins to be controlled by the activity of alpha particles, forward modes emerge and dominate the magnetic field profile at later times.

\begin{figure}
   \centering
   \includegraphics[width=8.5cm]{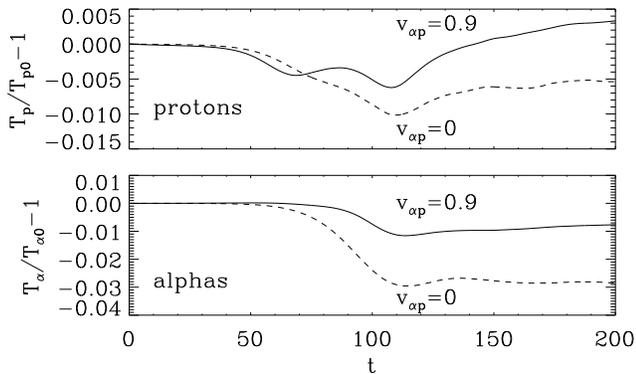} 
   \caption{Temporal variation of total ion temperatures, normalized to initial values, during fire hose instability for a simulation with (Run I, solid) and without (Run F, dashed) proton-alpha drift. Top panel shows the evolution of the total temperature of protons, bottom panel of alphas.
 }
   \label{fig_tot_t}
\end{figure}

Finally, we consider the evolution of the ion total temperatures $T_s$ during the instability. Fig.~\ref{fig_tot_t} shows the temporal profiles of $T_p$ (top) and $T_\alpha$ (bottom) normalized to the initial value, for Run F (dashed) and Run I (solid). When a relative drift between ion species is present, we observe a weaker cooling of the plasma during the fire hose activity. This is due to the fact that part of the kinetic energy associated to the drift motion is converted into thermal energy (heating) of each species by the instability. This partially compensates the loss of particle energy produced by the instability and results in a heating of each ion species.
The effect is stronger for protons; moreover, during the nonlinear phase of the instability in Run I, after $t\sim120$, we observe a further energization, leading to a net proton heating at expenses of the initial relative drift.

\section{Discussion}\label{discussion}
Properties of the alpha fire hose instability are found to be qualitatively similar to those of the instability generated by the protons. Fig.~\ref{fig_example} summarizes the global dynamics of the instability: the exponential growth of unstable modes extending over a finite range of k-vectors, the consequent reduction in anisotropy of particles by pitch-angle scattering and the associated deformation of the distribution function, developing characteristic non-thermal features.

Differences with respect to the lighter protons are however observed in the threshold of the instability and in the spectrum of unstable modes: fluctuations at larger wavelengths than in the case of protons are excited and the instability can be triggered at reasonably low betas for typical solar wind and space plasma conditions (see Figure \ref{fig_region}).
These properties can be qualitatively interpreted in terms of the resonant interactions that drive the fire hose instability in a kinetic regime.
As discussed by \cite{Verscharen_al_2013}, both protons and alpha particles become fire hose unstable for $v_{th, s}\gtrsim v_A$, where $v_{th,s}$ is the thermal speed of each species. Due to the difference in mass this translates into different conditions in terms of each species' beta - for a given plasma composition.
Despite the fact that dispersion relations in a plasma with multi-species anisotropies and drifts is really more complicated, this offers a simple argument to appreciate why in the case that fire hose is driven by alphas, and for solar wind-like abundances, the parallel fire hose unstable region can extend to $\beta_{\| \alpha}<1$ with a significant departure from the fluid non-resonant threshold, as shown in Figure \ref{fig_region}. Moreover, unlike the fluid non-resonant fire hose, the instability can be triggered also when the total beta of the plasma is well below unity (e.g., Run A).
Note that this can be relevant also for astrophysical and laboratory \citep{Carter_al_2013} systems characterized by small beta regimes, where we expect heavier ions to be possibly more efficient  in generating fire hose fluctuations, since protons are substantially fire hose stable for $\beta_{\| p}<1$.

We have found that growth rates of the alpha fire hose are sensitive to several parameters; the presence of a proton anisotropy - even not necessarily large enough to drive the protons unstable - leads to a significant increase in the alpha instability growth rate (Figure \ref{fig_gamma}), a trend which seems to be also confirmed by in situ observations \citep{Maruca_al_2012}.
The behavior of the plasma is in particular modified by the presence of a relative drift between species. This breaks the symmetry in the propagation of unstable modes, leading to a net flux of forward/backward waves, depending on the drift of the species that drives the instability.
This aspect is relevant for space plasmas, where differential streaming between the different components of the plasma is ubiquitously observed. As a consequence the parallel fire hose is expected to produce a preferential generation of inward or outward waves  depending on the species that drive the instability, thus offering a possible observational signature for the identification of fluctuations generated by specific components of the plasma.

On the other hand,  we have shown that the fire hose instability plays also a role in regulating/reducing the differential speed itself. This is known in the case of fire hose driven by protons and has been proven previously in numerical simulations \citep{Hellinger_travnicek_2006, Matteini_al_2011}, while we have found that in the case of an instability driven by the alphas, the effect can be even more significant.

The parallel fire hose and the (parallel) magnetosonic beam instability can be considered as limiting cases of one instability
driven by the same physical mechanism, the excess of kinetic/thermal energy along the magnetic field \citep[cf.,][]{Verscharen_al_2013b}.
Both instabilities lead to a significant reshaping of particle distributions.
While beam-type instabilities with isotropic populations act to reduce the relative streaming between the ions down to the a value which is typically $\sim v_A$, the presence of temperature anisotropies can lead to a reduction of the drift below that threshold \citep{Verscharen_al_2013}. Consistently with that, in our simulations, as a result of the fire hose activity, the alpha-proton drift is reduced well below the Alfvén speed ($v_{\alpha p}\sim0.6v_A$ in Figure \ref{fig_run2}), in agreement with solar wind observations \citep{Neugebauer_al_1996, Goldstein_al_2010}. This further suggests that an instability driven by temperature anisotropy of alphas can play a role in regulating the alpha-proton drift in the solar wind.
These results confirm that a complex interplay is possible between temperature anisotropy and relative drifts in space plasmas, and this needs more specific future investigations.

\section{Summary and conclusion}\label{conclusion}
In summary, we have investigated the properties of the parallel fire hose instability generated by an alpha particle pressure anisotropy.
We have shown, for the first time, simulations of the linear and nonlinear phases of the instability, and we have compared the results with linear and quasilinear theory predictions.
This study has led to several significant conclusions. First there is a good agreement in trends predicted by the linear and quasilinear theories and the fully nonlinear results obtained during the early phase of the simulations. Quasilinear calculations correctly predict the evolution of the temperatures as a response to the fluctuations generated by the instability, leading to transverse heating and parallel cooling for all ions. Remarkably, the same behavior is observed also in the case when the main species (protons) is stable and isotropic, as a consequence of the activity driven by the less abundant alphas.
This demonstrates that instabilities driven by secondary populations are also important for the global plasma energetics and that they can play a role in shaping the properties of the main ion population. Our simulations suggest that typical solar wind abundances ($\sim5\%$) are sufficient to trigger alpha fire hose growth rates that lead to dynamics with significant impact on the protons.

It is important to highlight that the role of instabilities driven by alpha temperature anisotropies is confirmed by direct solar wind observations \citep{Maruca_al_2012} demonstrating that, similarly to protons \citep{Hellinger_al_2006}, the histogram of observational counts in the plane ($\beta_{\| \alpha}, T_{\perp \alpha}/T_{\| \alpha}$) at 1AU shows apparent empirical boundaries that correspond well to the expected thresholds of kinetic instabilities in the same parameter space.
Moreover, when comparing the proton and alpha particle observations, the latter show a clear shift towards lower betas in their distribution (Figure 2 in \cite{Maruca_al_2012}). This scenario is consistent with the linear theory prediction (see Fig.~\ref{fig_region}), where the alpha unstable region extends to significantly smaller betas compared to both the case when the instability is driven by protons and the classical fluid threshold. As discussed in the previous section this can be explained in terms of the different effects of resonances that drive the instability for protons and for heavier alpha particles. Such a behavior is very well confirmed by our simulations, where it is clearly observed that  alphas and protons saturate and are constrained by fire hose instabilities in distinct regions of the parameter space ($\beta_{\|}, T_{\perp}/T_{\|}$), thus in very good agreement with the in situ observations.
Moreover, we highlight the fact that the solar wind different ions are bounded simultaneously in distinct regions of the parameter space, in good agreement with the distinct fire hose theoretical thresholds of both protons and alphas. This constitutes a further strong indication, together with previous studies \citep[e.g.,][]{Kasper_al_2002, Hellinger_al_2006, Marsch_al_2006,  Bale_al_2009, Matteini_al_2013}, that this instability is at work in limiting the temperature anisotropy of ion species in the solar wind.

Our simulations also show that as a result of the scattering by the unstable fluctuations, the fire hose instability driven by alphas influences the evolution of the alpha-proton drift. We suggest that this mechanism can play a role in regulating ion-beams in the solar wind expansion, since sub-Alfvénic drifts are often observed in the fast wind, a signature consistent with the results of our simulations.
Clearly more work is needed to test the role of the alpha fire hose instability in the expanding solar wind and its coupling with other processes and microinstabilities \citep{Hellinger_al_2015}.

Finally, note that our findings are primarily addressed to heliospheric investigations, where a direct comparison between in situ data and simulations may be achieved; for this reason we have adopted abundances and properties of the plasma that are typical of the heliospheric plasma.
However, several astrophysical environments, including galaxy clusters and accretion disks, are expected to be environments where pressure anisotropies play a role in regulating the plasma energetics \citep{Schekochihin_al_2005, Sharma_al_2006}. Despite some work about proton driven instability having been recently carried out \citep{Kunz_al_2014, Riquelme_al_2015}, we suggest that the contribution from other minor species of the plasma may be important as well.

\begin{acknowledgments}
This work was supported by UK Science and Technology Facilities Council grant ST/K001051/1.
The research leading to these results has received funding from the European Commission's Seventh Framework Programme (FP7/2007-2013) under the grant agreement SHOCK (project number 284515). PH acknowledges GACR grant 15-10057S. 
\end{acknowledgments}

\appendix
\section{Appendix: Symbols and units}\label{appendix}
Here $\perp$ and  $\parallel$ denote the directions with respect to the ambient magnetic field $\bf{B}_0$, $B_0=|\bf{B}_0|$ its magnitude, and
$\delta B$ denotes the magnitude of the fluctuating magnetic field; $v_\|$ and $v_\perp$ are velocity components parallel and perpendicular to 
$\bf{B}_0$, respectively; $t$ denotes the time.
Subscripts ${s}$ refer to different plasma species;
${p}$ stands for protons, ${\alpha}$ for alpha particles. 
$T_{\perp s}$ and $T_{\parallel s}$ are the perpendicular and parallel temperatures, respectively, and $A_s$ is the species temperature anisotropy $T_{\perp s}/T_{\| s}$.
$v_{th,s}=(k_B T_s/m_s)^{1/2}$ is the thermal velocity and 
$\beta_{\| s}=8\pi n_s k_B T_{\| s}/ B_0^2$ is the parallel plasma beta for a given plasma specie $s$ of density $n_s$ ($k_B$ is the
Boltzmann constant).

Through the paper the adopted units of space and time are the ion inertial length $c/\omega_{p}$ and 
the inverse proton cyclotron frequency $\Omega_{p}^{-1}$, respectively, where 
$\Omega_p=e B_o/m_p c$ and $\omega_p=(4\pi n e^2/m_p)^{1/2}$ is the proton plasma frequency. 
In these expressions $m_s$ and $e$ denote the mass and the charge, respectively. $v_{\alpha p}$ is the alpha-proton drift velocity and velocities are expressed in units of the Alfv\'en velocity $v_A=B_0/(4\pi n m_p)^{1/2}$.


\bibliographystyle{apj.bst}



\end{document}